\documentclass[twocolumn]{aastex6}
\usepackage{amsmath,amstext}
\usepackage[T1]{fontenc}
\usepackage[figure,figure*]{hypcap}
\usepackage{apjfonts}
\usepackage{graphicx}
\usepackage{amsmath}
\usepackage{dpfloat}
\usepackage{longtable}

\usepackage{xspace}
\newcommand{\nc}[2]{\newcommand{#1}{\ensuremath{#2}\xspace}}
\nc{\Kp}{ \textit{Kp} }
\nc{\SpecMatch}{\tt SpecMatch}
\nc{\isochrones}{\tt isochrones}
\nc{\TERRA}{\tt TERRA}
\nc{\teff}{ T_{\rm eff} }
\nc{\logg}{\log g}
\nc{\fe}{ \text{[Fe/H]}}

\usepackage{color}

\def\kms{\ifmmode{\rm km\thinspace s^{-1}}\else km\thinspace s$^{-1}$\fi}

\shortauthors{Winn et al.~2017}
\shorttitle{Absence of a metallicity effect for USP planets}

\begin{document}

%
\def\ltsima{$\; \buildrel < \over \sim \;$}
\def\lsim{\lower.5ex\hbox{\ltsima}}
\def\gtsima{$\; \buildrel > \over \sim \;$}
\def\gsim{\lower.5ex\hbox{\gtsima}}
%

\def\kms{\ifmmode{\rm km\thinspace s^{-1}}\else km\thinspace s$^{-1}$\fi}

\title{Absence of a metallicity effect for ultra-short-period planets\altaffilmark{1}}

\author{Joshua~N.~Winn\altaffilmark{2}}
\author{Roberto~Sanchis-Ojeda\altaffilmark{3}}
\author{Leslie~Rogers\altaffilmark{4}}
\author{Erik~A.~Petigura\altaffilmark{5}}
\author{Andrew~W.~Howard\altaffilmark{5}}
\author{Howard~Isaacson\altaffilmark{3}}
\author{Geoffrey~W.~Marcy\altaffilmark{3}}
\author{Kevin~C.~Schlaufman\altaffilmark{6}}
\author{Phillip Cargile\altaffilmark{7}}
\author{Leslie Hebb\altaffilmark{8}}

\altaffiltext{1}{Based on observations obtained at the W.\,M.\,Keck
  Observatory, which is operated jointly by the University of
  California and the California Institute of Technology. Keck time was
  granted by NASA, the University of California, the California
  Institute of Technology, and the University of Hawaii.}

\altaffiltext{2}{Department of Astrophysical Sciences, Princeton
  University, 4 Ivy Lane, Princeton, NJ 08540, USA}

\altaffiltext{3}{Department of Astronomy, University of California,
  Berkeley, CA 94720}

\altaffiltext{4}{Department of Astronomy \& Astrophysics, University of
  Chicago, 5640 South Ellis Avenue, Chicago, IL 60637, USA}

\altaffiltext{5}{Department of Astronomy, California Institute of
  Technology, Pasadena, CA 91125, USA}

\altaffiltext{6}{Department of Physics and Astronomy, Johns Hopkins
  University, Baltimore, MD 21218, USA}

\altaffiltext{7}{Harvard-Smithsonian Center for Astrophysics, 60
  Garden St, Cambridge, MA 02138, USA}

\altaffiltext{8}{Hobart and William Smith Colleges, Geneva, NY 14456, USA}

 \slugcomment{{\it Astronomical Journal}, in press}

\begin{abstract}

  Ultra-short-period (USP) planets are a newly recognized class of
  planets with periods shorter than one day and radii smaller than
  about 2~$R_\oplus$. It has been proposed that USP planets are the
  solid cores of hot Jupiters that lost their gaseous envelopes due to
  photo-evaporation or Roche lobe overflow. We test this hypothesis by
  asking whether USP planets are associated with metal-rich stars, as
  has long been observed for hot Jupiters.  We find the metallicity
  distributions of USP-planet and hot-Jupiter hosts to be
  significantly different ($p = 3\times 10^{-4}$), based on Keck
  spectroscopy of {\it Kepler} stars. Evidently, the sample of USP
  planets is not dominated by the evaporated cores of hot
  Jupiters. The metallicity distribution of stars with USP planets is
  indistinguishable from that of stars with short-period planets with
  sizes between 2--4~$R_\oplus$. Thus it remains possible that the USP
  planets are the solid cores of formerly gaseous planets smaller than
  Neptune.
      
\end{abstract}


\section{Introduction}

The discovery of planets with orbital periods shorter than one day,
and comparable in size to the Earth, has sparked discussion about
their origin and evolution. The first well-documented planets in this
category were CoRoT-7b \citep{Leger+2009}, Kepler-10b
\citep{Batalha+2011}, 55~Cnc~e \citep{DawsonFabrycky2010, Winn+2011,
  Demory+2011}, and Kepler-78b \citep{SanchisOjeda+2013}. A sample of
about 100 such planets was drawn together from {\it Kepler} data and
analyzed by \citet{SanchisOjeda+2014}. An independent {\it Kepler}
survey was performed by \citep{Jackson+2013}, and new examples have
since been discovered by \citet{Becker+2015}, \citet{Adams+2016}, and
\citet{Vanderburg+2016}.

Among the hypotheses for the origin of these ``ultra-short-period''
(USP) planets is that they are the exposed solid cores of hot Jupiters
that formed through core accretion. As circumstantial evidence for a
connection between USP planets and hot Jupiters, \citet{SanchisOjeda+2014}
and \citet{SteffenCoughlin2016} noted that these two categories of
planets are both found around $\approx$0.5\% of FGK stars. They also
found that USP planets are almost always smaller than 2~$R_\oplus$,
putting them in or near the size range for which planets are thought
to have a mainly rocky composition \citep{WeissMarcy2014,Rogers2015}.
They hypothesized that the most strongly irradiated hot Jupiters
eventually lose their gaseous envelopes due to photo-evaporation or
Roche lobe overflow \citep{Valsecchi+2014}.  This would leave behind a
nearly-naked core in a close-in orbit.  Proving this hypothesis to be
correct would confirm the core-accretion theory, and enable direct
measurements of the size and mass distribution of the rocky cores that
nucleate the growth of giant planets.

However, there are other possibilities for the origin of the USP
planets. They might represent the short-period extension of the
distribution of close-in rocky planets which either formed by core
accretion in their current orbits \citep{ChiangLaughlin2013}, or
migrated inwards from more distant orbits \citep{IdaLin2004,
  Schlaufman+2010, Terquem2014}. Another possibility is that the USP
planets are the exposed remnants not of hot Jupiters, but of smaller
gaseous planets with sizes between 2-4~$R_\oplus$
\citep{Lundkvist+2016,LeeChiang2017}.

\begin{figure*}[ht!]
 \begin{center}
 \leavevmode
 \epsscale{0.9}
 \plotone{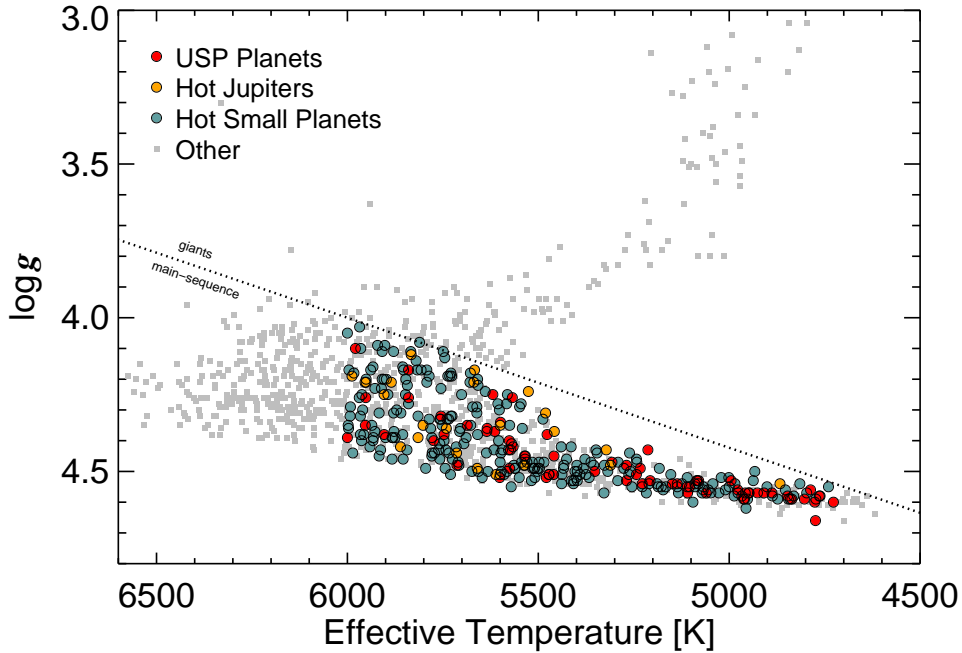}
 \end{center}
 \vspace{-0.25in}
  \caption{ {\bf Spectroscopic parameters of the stellar samples.}
    Stars below the dashed line were deemed main-sequence stars for
    the purpose of constructing our statistical samples, as described
    in \S\ref{sec:met}. Colored circles show the parameters of the
    sample stars. The smaller squares are the broader sample of
    stars in the California {\it Kepler} Survey
    \citep{Petigura+2017}.}
  \label{fig:teff_logg}
\end{figure*}

Here we test for a connection between USP planets and hot Jupiters by
comparing the metallicities of their host stars.  Stars that host
giant planets with orbital periods shorter than a few years have
systematically higher metallicities than randomly chosen stars in the
solar neighborhood \citep{Gonzalez1997, Santos+2004,
  FischerValenti2005}. In contrast, the host stars of smaller planets
show little if any association with high metallicity \citep{Udry+2006,
  Sousa+2011, SchlaufmanLaughlin2011, Buchhave+2012}.\footnote{We
  note, though, that these studies focused on stars near solar
  metallicity, and that \citet{Zhu+2016} have questioned some of the
  evidence. Furthermore, \citet{WangFischer2015} found a metallicity
  effect for small planets, though not as strong as for giant planets;
  and \citet{Adibekyan+2012} found that small-planet hosts tend to be
  higher in $\alpha$-elements even if they are relatively poor in iron
  (the traditional metallicity indicator).} If all USP planets are the
cores of former hot Jupiters, we should observe similar metallicity
distributions for the hosts of USP planets and hot Jupiters. If
instead the progenitors of USP planets are gaseous planets less
massive than hot Jupiters, or if they form in the same way as somewhat
longer-period planets, then the stars with USP planets should have a
metallicity distribution similar to that of short-period
sub-Neptunes. The metallicity distribution of {\it Kepler} planet
hosts has been investigated previously by \citet{Buchhave+2012,
  Mann+2013, Buchhave+2014, Dong+2014, Schlaufman2015,
  BuchhaveLatham2015, Guo+2016} and \citet{Mulders+2016}, but without
special attention to USP planets. Our study focuses on USP planets,
using the curated sample of \citet{SanchisOjeda+2014} and
metallicities from new high-resolution spectroscopy by
\citet{Petigura+2017}.  Section~\ref{sec:obs} describes our
observations and sample selection.  Section~\ref{sec:met} compares the
metallicity distributions of the host stars of hot Jupiters,
sub-Neptunes, and USP planets. Section~\ref{sec:mulders} compares our
results to those of \citet{Mulders+2016}.  Section~\ref{sec:concl}
provides some concluding remarks.

\section{Observations and Sample Selection}
\label{sec:obs}

\begin{figure*}
\begin{center}
 \epsscale{0.9}
 \plotone{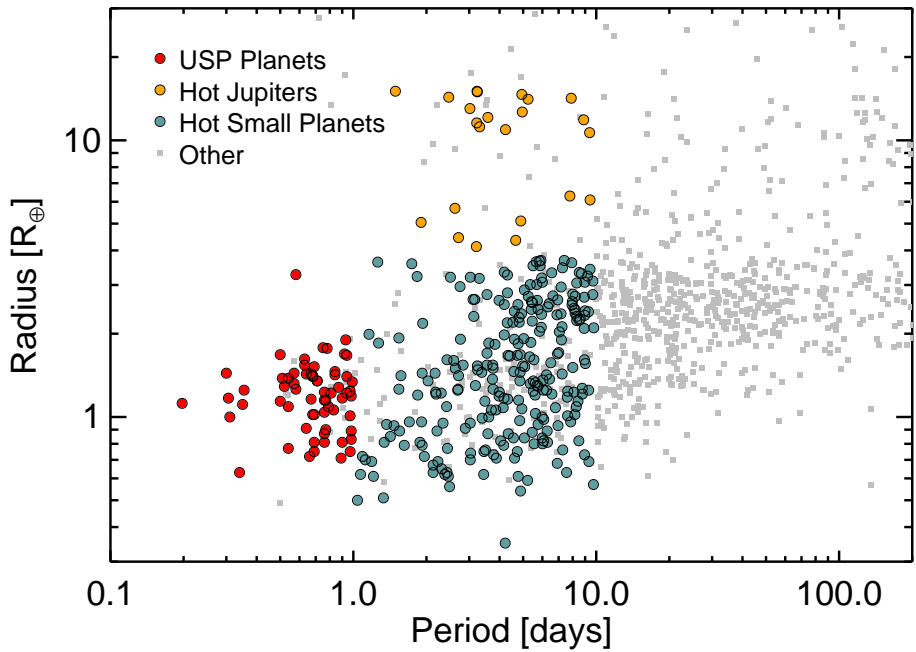}
  \end{center}
  \vspace{-0.25in}
  \caption{ {\bf Orbital period and planetary radius.}  The colored
    circles show our statistical samples; the smaller squares are for
    the broader sample of stars in the California {\it Kepler} Survey
    \citep{Petigura+2017}. }
  \label{fig:period_radius}
\end{figure*}

Sanchis-Ojeda et al.~(2014; hereafter, SO+14) presented a catalog of
USP planet candidates. We performed high-resolution optical
spectroscopy of 71 of the stars in this sample with the Keck~I
telescope and HIRES \citep{Vogt+1994}, as part of the larger
California Kepler Survey \citep[CKS;][]{Petigura+2017}. All the stars
brighter than $m_{\rm Kep} = 15.3$ were observed. Some fainter stars
were also observed, particularly those hosting the planets with the
shortest orbital periods. The spectra were collected from 2013~June to
2014~September. We used the standard California Planet Search setup,
but without the iodine cell, giving a typical spectral resolution of
$R=60,000$ over the wavelength range 0.36-0.80~$\mu$m. The exposure
times were typically 10 minutes, with a maximum exposure time of 20
minutes. For stars brighter than $m_{\rm Kep} = 14.3$, we achieved a
signal-to-noise ratio (SNR) of 40~pixel$^{-1}$ at 0.55~$\mu$m.  For
fainter stars, the SNR was between 20-40~pixel$^{-1}$.

The spectroscopic parameters of each star were determined with a
combination of \SpecMatch, a template-matching code; and a variant of
{\it Spectroscopy Made Easy}, a spectral synthesis code. Details are
provided by \citet{Petigura+2017}, who demonstrated a precision of
60~K in effective temperature, 0.07~dex in surface gravity, and
0.04~dex in [Fe/H].

For our study we omitted stars with $T_{\rm eff}<4700$~K. There are
severe discrepancies between the synthesized and observed spectra for
such cool stars, due to the onset of molecular absorption that is
poorly treated in the \citet{Coelho+2005} models. We also removed
KOI~2813 and KIC~5955905, for which the apparent transit signals have
been shown to be caused by binary stars rather than transiting
planets.\footnote{KOI 2813 was identified as a probable spectroscopic
  binary by \citet{Kolbl+2015}.  KIC~5955905 is a probable background
  binary, based on observations of large chromatic variations in the
  apparent transit depth (E.~Palle, private communication).}

The mass and radius of each star were determined by
\citet{Johnson+2017}, based on the comparison of the observed
spectroscopic parameters with those calculated with the Dartmouth
Stellar Evolution Program \citep{Dotter+2008}, using the \isochrones
code \citep{Morton+2016}\footnote{
  https://github.com/timothydmorton/isochrones~(version~1.0)}. The
inputs were \teff, \logg, and \fe, along with their associated
uncertainties. The code produces {\it a posteriori} distributions for
the stellar mass, radius, and age, by interpolating between the
available Dartmouth models.  The radii of the transiting planets were
then calculated from the stellar radii and the measured transit
depths.

\section{Metallicity distributions}
\label{sec:met}

We wanted to compare the metallicity distribution of the host stars of
USP planets, hot Jupiters, and non-giant planets with periods longer
than one day.  To construct the appropriate samples we drew on the
preceding results for the stars with USP planets, as well as the rest
of the stars in the California {\it Kepler} Survey
\citep{Petigura+2017}.  The larger sample includes about 1000 stars
selected from the list of {\it Kepler} Objects of Interest (KOI),
spanning a wide range of stellar types, planet sizes, and orbital
periods. The stars were selected for spectroscopy independently of
metallicity. Indeed, little information was available about the
metallicities prior to the observations.

We restricted our attention to main-sequence stars with effective
temperatures between 4700 and 6000~K, the range encompassing almost
all of the stars with USP planets.  We constructed three samples:
\begin{enumerate}

\item {\it USP Planets}:~Stars having a planet with orbital period shorter
  than 1~day, selected from SO+14 as described above. This sample has
  64 stars.

\item {\it Hot Jupiters}:~Stars with a planet larger than 4~$R_\oplus$
  and an orbital period shorter than 10~days. The somewhat arbitrary
  value of 4~$R_\oplus$ was chosen to match the value reported by
  \citet{Buchhave+2012} and \citet{Buchhave+2014} to distinguish
  different metallicity regimes. We omitted objects designated as
  ``False Positives'' by \citet{Twicken+2016} or
  \citet{Santerne+2016}.  We also omitted objects with inferred sizes
  larger than 20~$R_\oplus$ because experience has shown that in these
  cases the transit-like signal arises from a binary star rather than
  a transiting planet.  This sample has 23 stars.

\item {\it Hot Small Planets}:~Stars with planets smaller than
  4~$R_\oplus$ and orbital periods between 1-10~days, after omitting
  objects designated as ``False Positives''.  This sample has 246
  stars.

\end{enumerate}

\begin{figure*}
\begin{center}
 \epsscale{0.9}
 \plotone{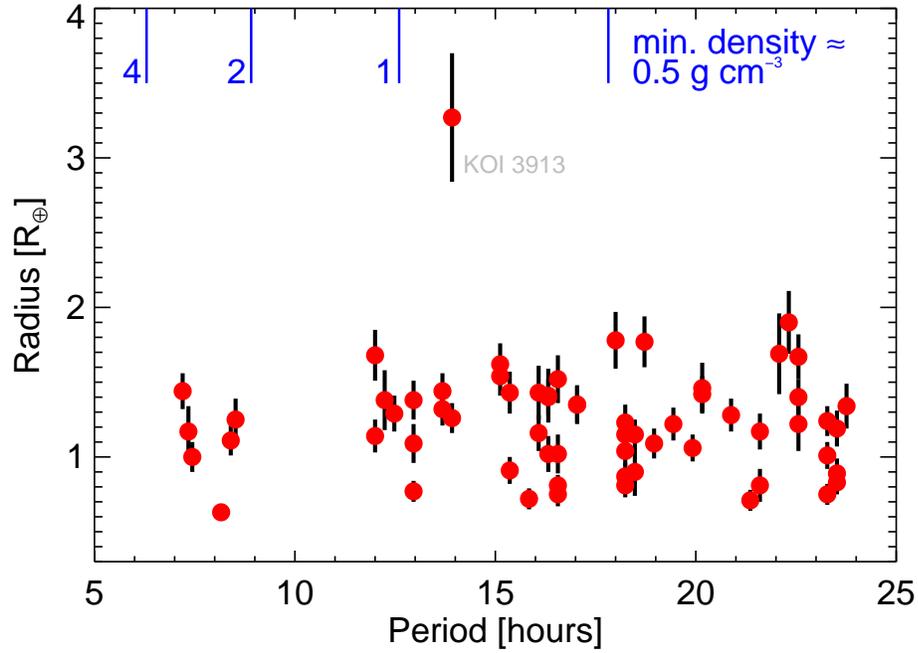}
 \end{center}
  \vspace{-0.25in}
 \caption{ {\bf Orbital period and planetary radius for USP planets.}
   Also marked are the Roche-limiting minimum periods for
   incompressible fluid bodies with mean densities of 0.5, 1.0, 2.0,
   and 4.0~g~cm$^{-3}$, using Eqn.~(2) of \citet{Rappaport+2013}. In
   reality, compression of the planetary interior may lower the
   minimum period by as much as $\approx$15\%.}
 \label{fig:period_radius_usp}
\end{figure*}

Tables~\ref{tbl:usp} and \ref{tbl:giant} give the pertinent properties
of the USP Planets and Hot Jupiters. Figure~\ref{fig:teff_logg} shows
the spectroscopic parameters \teff and \logg for the stars in each
sample. The dashed line is the boundary we used to identify
main-sequence stars; our samples were restricted to stars below this
line.  Figure~\ref{fig:period_radius} shows the period-radius
distribution of the planets hosted by the stars in each sample.  In
both figures, the small gray squares show the full sample of {\it
  Kepler} stars that were analyzed by \citet{Johnson+2017}.

Figure~\ref{fig:period_radius_usp} focuses exclusively on the USP
Planets. All but one of the USP planets have sizes
$\lsim$2~$R_\oplus$, even though no selection was made based on planet
size.  Thus, we confirm the finding of SO+14 that USP planets are
almost always smaller than 2~$R_\oplus$.  We find no major differences
between our newly-determined radius distribution and the distribution
presented by SO+14 except that the new estimates of planetary radii
have smaller uncertainties, and one of the outliers with size
$>$2~$R_\oplus$ does not appear in the new sample. The single
remaining USP planet with $R>2~R_{\oplus}$ is KOI~3913, a remarkable
case which deserves further observations.

\begin{figure*}[ht!]
\begin{center}
 \leavevmode
 \epsscale{0.9}
 \plotone{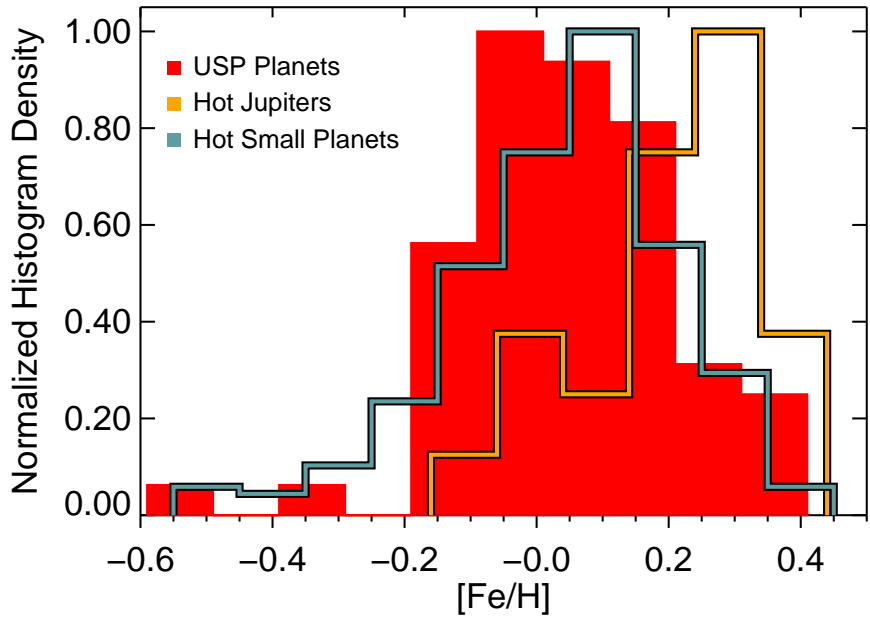}
\end{center}
\caption{ {\bf Metallicity distributions} of the three statistical
  samples.  The hot-Jupiter hosts have a different metallicity
  distribution (more weighted toward high metallicity) than the hosts of USP planets,
  and the hosts of planets smaller than Neptune with periods 1-10~days.}
  \label{fig:metallicity}
\end{figure*}

Figure~\ref{fig:metallicity} shows the distribution of \fe for the
stars in each sample. Even at a glance, the Hot Jupiters are seen to
be weighted toward higher \fe than either the USP Planets or the Hot
Small Planets. The distributions for the USP and Hot Small Planets
appear similar to one another. To quantify these impressions we
performed two-sample Kolmogorov-Smirnov tests, which estimate the
probability $p$ that two samples are drawn from the same distribution.
The results, given in Table~\ref{tbl:pvalues}, indicate that the USP
Planets and Hot Jupiters are very unlikely to be drawn from the same
distribution, while the USP Planets and the Hot Smaller Planets have
distributions that are indistinguishable with the current data.

\section{Upper bound on hot-Jupiter fraction}

Evidently the stars with USP planets have a different metallicity
distribution than stars with hot Jupiters.  We placed an upper bound
on the fraction $f$ of members in the USP Planet sample that could
have been drawn from the same distribution as the Hot Jupiter sample,
using a Monte Carlo technique. We considered the range of $f$ from
zero to unity. For each choice of $f$, we constructed a sample of 64
metallicities, matching the actual USP sample size. We randomly drew
(with replacement) $N=[64f]$ values from the Hot Jupiter sample and
$64-N$ values from the USP sample, where $[x]$ indicates rounding
to the nearest integer. We added Gaussian errors to each metallicity
with a standard deviation of 0.04~dex. We then computed the
probability $p$ that the simulated sample was drawn from the same
underlying distribution as the Hot Jupiter sample, using a two-sided
Kolmogorov-Smirnov test. We repeated this procedure $10^3$ times and
recorded the mean $p$ value.

For high values of $f$, the simulated sample is drawn entirely from the
USP Planets and the $p$-values are $\sim$10$^{-3}$ as described in the
previous section. For values of $f$ approaching unity, the $p$-values
are $\sim$1 because the Hot Jupiter sample is being compared with
itself. To determine an upper bound on $f$, we sought the value for
which $p=0.0455$, corresponding to a nominal 2$\sigma$ level of
confidence. The result is $f<0.36$, implying that no more than about
half of the metallicities of the USP host stars could have been drawn
from the same metallicity distribution as the Hot Jupiter hosts.

\section{Comparison with Mulders et al.~(2016)}
\label{sec:mulders}

\citet{Mulders+2016} studied the relationship between orbital period
and stellar metallicity for a sample of 665 {\it Kepler} planet
candidates. They found the mean metallicity of stars with planets
shorter than 10 days to be higher than for stars with longer-period
planets.  This was true for all planet sizes, with the strongest
effect ($+0.25\pm 0.07$~dex) seen for the smallest planets
($<$1.7~$R_\oplus$).

Our study is concerned exclusively with planets with $P<10$~days. To
compare our data with theirs, we note that their Fig.~2 shows the mean
metallicity to be nearly constant for periods ranging from 0.6-5~days,
before decreasing by $\approx$0.1~dex from 5-10~days.  Our data does
not display such a period dependence: the hosts of planets with
periods $<$5~days and 5-10~days have the same mean metallicity to
within 0.02~dex. A broader comparison between the CKS metallicity
scale and that of \citet{Mulders+2016} also shows significant
differences, which will be examined as part of a forthcoming CKS paper
led by E.~Petigura.

We also note that our sample of ultra-short-period planets differs
from that of \citet{Mulders+2016}. Our sample has 65 planets, while
their sample has eight planets, three of which (KOI~2813, 2717 and
3204) have host stars with effective temperatures outside the range of
our study (4700-6000~K).

\section{Conclusions}
\label{sec:concl}

The metallicity distribution of the host stars of USP planets does not
resemble the metallicity distribution of the host stars of hot
Jupiters. In particular, the stars with USP planets show no evidence
for an association with high metallicity, unlike stars with hot
Jupiters.  The USP hosts have a mean metallicity near the Sun's value,
and similar to that of the general planet-hosting population of {\it
  Kepler} stars.

This basic result is also obtained if we make some small changes to
our sample definitions.  If we require ``Hot Jupiters'' to have radii
larger than 7~$R_\oplus$, instead of 4~$R_\oplus$, then the sample
size decreases from 23 to 15.  The metallicity distribution of their
host stars remains distinguishable from that of the stars with USP
planets, though with reduced statistical significance ($p=0.01$).  We
also tried requiring the ``Hot Small Planets'' to have radii between
2-4~$R_\oplus$, i.e., we omitted the smaller planets that are more
likely to be rocky. This is a more direct test of an evolutionary
connection between USP planets and gas-rich planets at slightly longer
periods. This reduces the sample size from 246 to 82.  When tested
against the metallicity distribution of the stars with USP planets,
the $p$-value changes from 0.39 to 0.10, which is still too large to
be considered evidence for a significant difference.

We interpret these results as an argument against any theory in which
most of the USP planets are descended from hot Jupiters.  In such a
theory, the stars that are currently observed to have USP planets were
once hosts to hot Jupiters, and their metallicity distribution should
be the same as those stars currently observed to have hot Jupiters.
The only way we see to escape this conclusion---which seems
unlikely---is to hypothesize that the process that converts hot
Jupiters into USP planets also systematically lowers the metallicity
of the host star by $\Delta {\rm [Fe/H]} \approx -0.15$, so as to
match the metallicity distribution of the hosts of smaller {\it
  Kepler} planets.

The possibility that USP planets represent the solid cores of
erstwhile hot Jupiters had already been deemed unlikely on theoretical
grounds, because of the difficulty of removing such a massive gaseous
atmosphere. \citet{MurrayClay+2009} modeled the wind launched from a
gaseous planet by a star's high-energy radiation, and found it
difficult to erode the entire atmosphere of a hot Jupiter.  Had we
found a strong metallicity enhancement for the hosts of USP planets,
this theoretical conclusion would have been called into question.

It remains plausible that the progenitors of USP planets are
Neptune-sized or smaller planets with gaseous atmospheres.  This is
also compatible with the tendency of USP planets to have sub-Neptune
companions in somewhat wider orbits
\citep{SanchisOjeda+2014,Adams+2017}. Multiple theoretical studies
have shown that it is possible to lose most of the gas from a
low-density planet smaller than Neptune \citep{HoweBurrows2015,
  Lopez2016, Jackson+2017, SivanGinzburg2017}. Also consistent with
this picture is the recent discovery by \citet{Fulton+2017} that
relatively few {\it Kepler} planets have sizes between
1.5-2~$R_\oplus$. The missing planets in this size range might have
been gas-rich sub-Neptunes whose atmospheres were stripped.

Ultra-short-period planets remain an attractive subject for future
work to understand their origin, occurrence rate, radius distribution,
and the dependence of all these quantities on the properties of the
host star. The current sample of $\sim$100 stars with USP planets have
apparent magnitudes that are generally too faint for precise Doppler
monitoring, observations of the Rossiter-McLaughlin effect, and
detections of occultations or transmission effects. The {\it TESS}
mission \citep{Ricker+2015} will help to remedy this problem by
searching a similar number of stars as the {\it Kepler} mission, but
brighter by several magnitudes.

\acknowledgements We thank the anonymous referee for a prompt and
helpful report. We thank Simon Albrecht, Eugene Chiang, Brian Jackson,
Eve Lee, Peter McCullough, Saul Rappaport, Amaury Triaud and Francesca
Valsecchi for helpful discussions. We also thank Enric Palle for
sharing his results for KIC~5955905. J.N.W.\ acknowledges the support
from a NASA Keck PI Data Award, administered by the NASA Exoplanet
Science Institute. A.W.H.\ acknowledges the support from NASA grant
NNX12AJ23G. Data presented herein were obtained at the W.M.\ Keck
Observatory from telescope time allocated to the National Aeronautics
and Space Administration through the agency's scientific partnership
with the California Institute of Technology and the University of
California. The Observatory was made possible by the generous
financial support of the W.M.\ Keck Foundation. The authors
acknowledge the very significant cultural role and reverence that the
summit of Mauna Kea has always had within the indigenous Hawaiian
community. We are most fortunate to have the opportunity to conduct
observations from this mountain.

\bibliographystyle{yahapj}
\bibliography{references}

\begin{thebibliography}{}
\providecommand\natexlab[1]{#1}
\providecommand\JournalTitle[1]{#1}

\bibitem[{{Adams} {et~al.}(2016){Adams}, {Jackson}, \& {Endl}}]{Adams+2016}
{Adams}, E.~R., {Jackson}, B., \& {Endl}, M. 2016,
  \href{http://dx.doi.org/10.3847/0004-6256/152/2/47}{\JournalTitle{\aj}, 152,
  47}

\bibitem[{{Adams} {et~al.}(2017){Adams}, {Jackson}, {Endl}, {Cochran},
  {MacQueen}, {Duev}, {Jensen-Clem}, {Salama}, {Ziegler}, {Baranec},
  {Kulkarni}, {Law}, \& {Riddle}}]{Adams+2017}
{Adams}, E.~R., {Jackson}, B., {Endl}, M., {et~al.} 2017,
  \href{http://dx.doi.org/10.3847/1538-3881/153/2/82}{\JournalTitle{\aj}, 153,
  82}

\bibitem[{{Adibekyan} {et~al.}(2012){Adibekyan}, {Delgado Mena}, {Sousa},
  {Santos}, {Israelian}, {Gonz{\'a}lez Hern{\'a}ndez}, {Mayor}, \&
  {Hakobyan}}]{Adibekyan+2012}
{Adibekyan}, V.~Z., {Delgado Mena}, E., {Sousa}, S.~G., {et~al.} 2012,
  \href{http://dx.doi.org/10.1051/0004-6361/201220167}{\JournalTitle{\aap},
  547, A36}

\bibitem[{{Batalha} {et~al.}(2011){Batalha}, {Borucki}, {Bryson}, {Buchhave},
  {Caldwell}, {Christensen-Dalsgaard}, {Ciardi}, {Dunham}, {Fressin},
  {Gautier}, {Gilliland}, {Haas}, {Howell}, {Jenkins}, {Kjeldsen}, {Koch},
  {Latham}, {Lissauer}, {Marcy}, {Rowe}, {Sasselov}, {Seager}, {Steffen},
  {Torres}, {Basri}, {Brown}, {Charbonneau}, {Christiansen}, {Clarke},
  {Cochran}, {Dupree}, {Fabrycky}, {Fischer}, {Ford}, {Fortney}, {Girouard},
  {Holman}, {Johnson}, {Isaacson}, {Klaus}, {Machalek}, {Moorehead},
  {Morehead}, {Ragozzine}, {Tenenbaum}, {Twicken}, {Quinn}, {VanCleve},
  {Walkowicz}, {Welsh}, {Devore}, \& {Gould}}]{Batalha+2011}
{Batalha}, N.~M., {Borucki}, W.~J., {Bryson}, S.~T., {et~al.} 2011,
  \href{http://dx.doi.org/10.1088/0004-637X/729/1/27}{\JournalTitle{\apj}, 729,
  27}

\bibitem[{{Becker} {et~al.}(2015){Becker}, {Vanderburg}, {Adams}, {Rappaport},
  \& {Schwengeler}}]{Becker+2015}
{Becker}, J.~C., {Vanderburg}, A., {Adams}, F.~C., {Rappaport}, S.~A., \&
  {Schwengeler}, H.~M. 2015,
  \href{http://dx.doi.org/10.1088/2041-8205/812/2/L18}{\JournalTitle{\apjl},
  812, L18}

\bibitem[{{Buchhave} \& {Latham}(2015)}]{BuchhaveLatham2015}
{Buchhave}, L.~A., \& {Latham}, D.~W. 2015,
  \href{http://dx.doi.org/10.1088/0004-637X/808/2/187}{\JournalTitle{\apj},
  808, 187}

\bibitem[{{Buchhave} {et~al.}(2012){Buchhave}, {Latham}, {Johansen},
  {Bizzarro}, {Torres}, {Rowe}, {Batalha}, {Borucki}, {Brugamyer}, {Caldwell},
  {Bryson}, {Ciardi}, {Cochran}, {Endl}, {Esquerdo}, {Ford}, {Geary},
  {Gilliland}, {Hansen}, {Isaacson}, {Laird}, {Lucas}, {Marcy}, {Morse},
  {Robertson}, {Shporer}, {Stefanik}, {Still}, \& {Quinn}}]{Buchhave+2012}
{Buchhave}, L.~A., {Latham}, D.~W., {Johansen}, A., {et~al.} 2012,
  \href{http://dx.doi.org/10.1038/nature11121}{\JournalTitle{\nat}, 486, 375}

\bibitem[{{Buchhave} {et~al.}(2014){Buchhave}, {Bizzarro}, {Latham},
  {Sasselov}, {Cochran}, {Endl}, {Isaacson}, {Juncher}, \&
  {Marcy}}]{Buchhave+2014}
{Buchhave}, L.~A., {Bizzarro}, M., {Latham}, D.~W., {et~al.} 2014,
  \href{http://dx.doi.org/10.1038/nature13254}{\JournalTitle{\nat}, 509, 593}

\bibitem[{{Chiang} \& {Laughlin}(2013)}]{ChiangLaughlin2013}
{Chiang}, E., \& {Laughlin}, G. 2013,
  \href{http://dx.doi.org/10.1093/mnras/stt424}{\JournalTitle{\mnras}, 431,
  3444}

\bibitem[{{Coelho} {et~al.}(2005){Coelho}, {Barbuy}, {Mel{\'e}ndez},
  {Schiavon}, \& {Castilho}}]{Coelho+2005}
{Coelho}, P., {Barbuy}, B., {Mel{\'e}ndez}, J., {Schiavon}, R.~P., \&
  {Castilho}, B.~V. 2005,
  \href{http://dx.doi.org/10.1051/0004-6361:20053511}{\JournalTitle{\aap}, 443,
  735}

\bibitem[{{Dawson} \& {Fabrycky}(2010)}]{DawsonFabrycky2010}
{Dawson}, R.~I., \& {Fabrycky}, D.~C. 2010,
  \href{http://dx.doi.org/10.1088/0004-637X/722/1/937}{\JournalTitle{\apj},
  722, 937}

\bibitem[{{Demory} {et~al.}(2011){Demory}, {Gillon}, {Deming}, {Valencia},
  {Seager}, {Benneke}, {Lovis}, {Cubillos}, {Harrington}, {Stevenson}, {Mayor},
  {Pepe}, {Queloz}, {S{\'e}gransan}, \& {Udry}}]{Demory+2011}
{Demory}, B.-O., {Gillon}, M., {Deming}, D., {et~al.} 2011,
  \href{http://dx.doi.org/10.1051/0004-6361/201117178}{\JournalTitle{\aap},
  533, A114}

\bibitem[{{Dong} {et~al.}(2014){Dong}, {Zheng}, {Zhu}, {De Cat}, {Fu}, {Yang},
  {Zhang}, {Jin}, \& {Zhang}}]{Dong+2014}
{Dong}, S., {Zheng}, Z., {Zhu}, Z., {et~al.} 2014,
  \href{http://dx.doi.org/10.1088/2041-8205/789/1/L3}{\JournalTitle{\apjl},
  789, L3}

\bibitem[{{Dotter} {et~al.}(2008){Dotter}, {Chaboyer}, {Jevremovi{\'c}},
  {Kostov}, {Baron}, \& {Ferguson}}]{Dotter+2008}
{Dotter}, A., {Chaboyer}, B., {Jevremovi{\'c}}, D., {et~al.} 2008,
  \href{http://dx.doi.org/10.1086/589654}{\JournalTitle{\apjs}, 178, 89}

\bibitem[{{Fischer} \& {Valenti}(2005)}]{FischerValenti2005}
{Fischer}, D.~A., \& {Valenti}, J. 2005,
  \href{http://dx.doi.org/10.1086/428383}{\JournalTitle{\apj}, 622, 1102}

\bibitem[{{Fulton} {et~al.}(2017){Fulton}, {Petigura}, {Howard}, {Isaacson},
  {Marcy}, {Cargile}, {Hebb}, {Weiss}, {Johnson}, {Morton}, {Sinukoff},
  {Crossfield}, \& {Hirsch}}]{Fulton+2017}
{Fulton}, B.~J., {Petigura}, E.~A., {Howard}, A.~W., {et~al.} 2017,
  \JournalTitle{ArXiv e-prints},
  \href{http://arxiv.org/abs/1703.10375}{{\sffamily arXiv:1703.10375
  [astro-ph.EP]}}

\bibitem[{{Ginzburg} \& {Sari}(2016)}]{SivanGinzburg2017}
{Ginzburg}, S., \& {Sari}, R. 2016, \JournalTitle{ArXiv e-prints},
  \href{http://arxiv.org/abs/1611.09373}{{\sffamily arXiv:1611.09373
  [astro-ph.EP]}}

\bibitem[{{Gonzalez}(1997)}]{Gonzalez1997}
{Gonzalez}, G. 1997,
  \href{http://dx.doi.org/10.1093/mnras/285.2.403}{\JournalTitle{\mnras}, 285,
  403}

\bibitem[{{Guo} {et~al.}(2016){Guo}, {Johnson}, {Mann}, {Kraus}, {Curtis}, \&
  {Latham}}]{Guo+2016}
{Guo}, X., {Johnson}, J.~A., {Mann}, A.~W., {et~al.} 2016, \JournalTitle{ArXiv
  e-prints}, \href{http://arxiv.org/abs/1612.01616}{{\sffamily arXiv:1612.01616
  [astro-ph.SR]}}

\bibitem[{{Howe} \& {Burrows}(2015)}]{HoweBurrows2015}
{Howe}, A.~R., \& {Burrows}, A. 2015,
  \href{http://dx.doi.org/10.1088/0004-637X/808/2/150}{\JournalTitle{\apj},
  808, 150}

\bibitem[{{Ida} \& {Lin}(2004)}]{IdaLin2004}
{Ida}, S., \& {Lin}, D.~N.~C. 2004,
  \href{http://dx.doi.org/10.1086/424830}{\JournalTitle{\apj}, 616, 567}

\bibitem[{{Jackson} {et~al.}(2017){Jackson}, {Arras}, {Penev}, {Peacock}, \&
  {Marchant}}]{Jackson+2017}
{Jackson}, B., {Arras}, P., {Penev}, K., {Peacock}, S., \& {Marchant}, P. 2017,
  \href{http://dx.doi.org/10.3847/1538-4357/835/2/145}{\JournalTitle{\apj},
  835, 145}

\bibitem[{{Jackson} {et~al.}(2013){Jackson}, {Stark}, {Adams}, {Chambers}, \&
  {Deming}}]{Jackson+2013}
{Jackson}, B., {Stark}, C.~C., {Adams}, E.~R., {Chambers}, J., \& {Deming}, D.
  2013,
  \href{http://dx.doi.org/10.1088/0004-637X/779/2/165}{\JournalTitle{\apj},
  779, 165}

\bibitem[{{Johnson} {et~al.}(2017){Johnson}, {Petigura}, {Fulton}, {Marcy},
  {Howard}, {Isaacson}, {Hebb}, {Cargile}, {Morton}, {Weiss}, {Winn}, {Rogers},
  {Sinukoff}, \& {Hirsch}}]{Johnson+2017}
{Johnson}, J.~A., {Petigura}, E.~A., {Fulton}, B.~J., {et~al.} 2017,
  \JournalTitle{ArXiv e-prints},
  \href{http://arxiv.org/abs/1703.10402}{{\sffamily arXiv:1703.10402
  [astro-ph.EP]}}

\bibitem[{{Kolbl} {et~al.}(2015){Kolbl}, {Marcy}, {Isaacson}, \&
  {Howard}}]{Kolbl+2015}
{Kolbl}, R., {Marcy}, G.~W., {Isaacson}, H., \& {Howard}, A.~W. 2015,
  \href{http://dx.doi.org/10.1088/0004-6256/149/1/18}{\JournalTitle{\aj}, 149,
  18}

\bibitem[{{Lee} \& {Chiang}(2017)}]{LeeChiang2017}
{Lee}, E.~J., \& {Chiang}, E. 2017, \JournalTitle{ArXiv e-prints},
  \href{http://arxiv.org/abs/1702.08461}{{\sffamily arXiv:1702.08461
  [astro-ph.EP]}}

\bibitem[{{L{\'e}ger} {et~al.}(2009){L{\'e}ger}, {Rouan}, {Schneider}, {Barge},
  {Fridlund}, {Samuel}, {Ollivier}, {Guenther}, {Deleuil}, {Deeg}, {Auvergne},
  {Alonso}, {Aigrain}, {Alapini}, {Almenara}, {Baglin}, {Barbieri}, {Bruntt},
  {Bord{\'e}}, {Bouchy}, {Cabrera}, {Catala}, {Carone}, {Carpano}, {Csizmadia},
  {Dvorak}, {Erikson}, {Ferraz-Mello}, {Foing}, {Fressin}, {Gandolfi},
  {Gillon}, {Gondoin}, {Grasset}, {Guillot}, {Hatzes}, {H{\'e}brard}, {Jorda},
  {Lammer}, {Llebaria}, {Loeillet}, {Mayor}, {Mazeh}, {Moutou}, {P{\"a}tzold},
  {Pont}, {Queloz}, {Rauer}, {Renner}, {Samadi}, {Shporer}, {Sotin}, {Tingley},
  {Wuchterl}, {Adda}, {Agogu}, {Appourchaux}, {Ballans}, {Baron}, {Beaufort},
  {Bellenger}, {Berlin}, {Bernardi}, {Blouin}, {Baudin}, {Bodin}, {Boisnard},
  {Boit}, {Bonneau}, {Borzeix}, {Briet}, {Buey}, {Butler}, {Cailleau},
  {Cautain}, {Chabaud}, {Chaintreuil}, {Chiavassa}, {Costes}, {Cuna Parrho},
  {de Oliveira Fialho}, {Decaudin}, {Defise}, {Djalal}, {Epstein}, {Exil},
  {Faur{\'e}}, {Fenouillet}, {Gaboriaud}, {Gallic}, {Gamet}, {Gavalda},
  {Grolleau}, {Gruneisen}, {Gueguen}, {Guis}, {Guivarc'h}, {Guterman},
  {Hallouard}, {Hasiba}, {Heuripeau}, {Huntzinger}, {Hustaix}, {Imad},
  {Imbert}, {Johlander}, {Jouret}, {Journoud}, {Karioty}, {Kerjean},
  {Lafaille}, {Lafond}, {Lam-Trong}, {Landiech}, {Lapeyrere}, {Larqu{\'e}},
  {Laudet}, {Lautier}, {Lecann}, {Lefevre}, {Leruyet}, {Levacher}, {Magnan},
  {Mazy}, {Mertens}, {Mesnager}, {Meunier}, {Michel}, {Monjoin}, {Naudet},
  {Nguyen-Kim}, {Orcesi}, {Ottacher}, {Perez}, {Peter}, {Plasson}, {Plesseria},
  {Pontet}, {Pradines}, {Quentin}, {Reynaud}, {Rolland}, {Rollenhagen},
  {Romagnan}, {Russ}, {Schmidt}, {Schwartz}, {Sebbag}, {Sedes}, {Smit},
  {Steller}, {Sunter}, {Surace}, {Tello}, {Tiph{\`e}ne}, {Toulouse}, {Ulmer},
  {Vandermarcq}, {Vergnault}, {Vuillemin}, \& {Zanatta}}]{Leger+2009}
{L{\'e}ger}, A., {Rouan}, D., {Schneider}, J., {et~al.} 2009,
  \href{http://dx.doi.org/10.1051/0004-6361/200911933}{\JournalTitle{\aap},
  506, 287}

\bibitem[{{Lopez}(2016)}]{Lopez2016}
{Lopez}, E.~D. 2016, \JournalTitle{ArXiv e-prints},
  \href{http://arxiv.org/abs/1610.01170}{{\sffamily arXiv:1610.01170
  [astro-ph.EP]}}

\bibitem[{{Lundkvist} {et~al.}(2016){Lundkvist}, {Kjeldsen}, {Albrecht},
  {Davies}, {Basu}, {Huber}, {Justesen}, {Karoff}, {Silva Aguirre}, {van
  Eylen}, {Vang}, {Arentoft}, {Barclay}, {Bedding}, {Campante}, {Chaplin},
  {Christensen-Dalsgaard}, {Elsworth}, {Gilliland}, {Handberg}, {Hekker},
  {Kawaler}, {Lund}, {Metcalfe}, {Miglio}, {Rowe}, {Stello}, {Tingley}, \&
  {White}}]{Lundkvist+2016}
{Lundkvist}, M.~S., {Kjeldsen}, H., {Albrecht}, S., {et~al.} 2016,
  \href{http://dx.doi.org/10.1038/ncomms11201}{\JournalTitle{Nature
  Communications}, 7, 11201}

\bibitem[{{Mann} {et~al.}(2013){Mann}, {Gaidos}, {Kraus}, \&
  {Hilton}}]{Mann+2013}
{Mann}, A.~W., {Gaidos}, E., {Kraus}, A., \& {Hilton}, E.~J. 2013,
  \href{http://dx.doi.org/10.1088/0004-637X/770/1/43}{\JournalTitle{\apj}, 770,
  43}

\bibitem[{{Morton} {et~al.}(2016){Morton}, {Bryson}, {Coughlin}, {Rowe},
  {Ravichandran}, {Petigura}, {Haas}, \& {Batalha}}]{Morton+2016}
{Morton}, T.~D., {Bryson}, S.~T., {Coughlin}, J.~L., {et~al.} 2016,
  \href{http://dx.doi.org/10.3847/0004-637X/822/2/86}{\JournalTitle{\apj}, 822,
  86}

\bibitem[{{Mulders} {et~al.}(2016){Mulders}, {Pascucci}, {Apai}, {Frasca}, \&
  {Molenda-{\.Z}akowicz}}]{Mulders+2016}
{Mulders}, G.~D., {Pascucci}, I., {Apai}, D., {Frasca}, A., \&
  {Molenda-{\.Z}akowicz}, J. 2016,
  \href{http://dx.doi.org/10.3847/0004-6256/152/6/187}{\JournalTitle{\aj}, 152,
  187}

\bibitem[{{Murray-Clay} {et~al.}(2009){Murray-Clay}, {Chiang}, \&
  {Murray}}]{MurrayClay+2009}
{Murray-Clay}, R.~A., {Chiang}, E.~I., \& {Murray}, N. 2009,
  \href{http://dx.doi.org/10.1088/0004-637X/693/1/23}{\JournalTitle{\apj}, 693,
  23}

\bibitem[{{Petigura} {et~al.}(2017){Petigura}, {Howard}, {Marcy}, {Johnson},
  {Isaacson}, {Cargile}, {Hebb}, {Fulton}, {Weiss}, {Morton}, {Winn}, {Rogers},
  {Sinukoff}, {Hirsch}, \& {Crossfield}}]{Petigura+2017}
{Petigura}, E.~A., {Howard}, A.~W., {Marcy}, G.~W., {et~al.} 2017,
  \JournalTitle{ArXiv e-prints},
  \href{http://arxiv.org/abs/1703.10400}{{\sffamily arXiv:1703.10400
  [astro-ph.EP]}}

\bibitem[{{Rappaport} {et~al.}(2013){Rappaport}, {Sanchis-Ojeda}, {Rogers},
  {Levine}, \& {Winn}}]{Rappaport+2013}
{Rappaport}, S., {Sanchis-Ojeda}, R., {Rogers}, L.~A., {Levine}, A., \& {Winn},
  J.~N. 2013,
  \href{http://dx.doi.org/10.1088/2041-8205/773/1/L15}{\JournalTitle{\apjl},
  773, L15}

\bibitem[{{Ricker} {et~al.}(2015){Ricker}, {Winn}, {Vanderspek}, {Latham},
  {Bakos}, {Bean}, {Berta-Thompson}, {Brown}, {Buchhave}, {Butler}, {Butler},
  {Chaplin}, {Charbonneau}, {Christensen-Dalsgaard}, {Clampin}, {Deming},
  {Doty}, {De Lee}, {Dressing}, {Dunham}, {Endl}, {Fressin}, {Ge}, {Henning},
  {Holman}, {Howard}, {Ida}, {Jenkins}, {Jernigan}, {Johnson}, {Kaltenegger},
  {Kawai}, {Kjeldsen}, {Laughlin}, {Levine}, {Lin}, {Lissauer}, {MacQueen},
  {Marcy}, {McCullough}, {Morton}, {Narita}, {Paegert}, {Palle}, {Pepe},
  {Pepper}, {Quirrenbach}, {Rinehart}, {Sasselov}, {Sato}, {Seager},
  {Sozzetti}, {Stassun}, {Sullivan}, {Szentgyorgyi}, {Torres}, {Udry}, \&
  {Villasenor}}]{Ricker+2015}
{Ricker}, G.~R., {Winn}, J.~N., {Vanderspek}, R., {et~al.} 2015,
  \href{http://dx.doi.org/10.1117/1.JATIS.1.1.014003}{\JournalTitle{Journal of
  Astronomical Telescopes, Instruments, and Systems}, 1, 014003}

\bibitem[{{Rogers}(2015)}]{Rogers2015}
{Rogers}, L.~A. 2015,
  \href{http://dx.doi.org/10.1088/0004-637X/801/1/41}{\JournalTitle{\apj}, 801,
  41}

\bibitem[{{Sanchis-Ojeda} {et~al.}(2014){Sanchis-Ojeda}, {Rappaport}, {Winn},
  {Kotson}, {Levine}, \& {El Mellah}}]{SanchisOjeda+2014}
{Sanchis-Ojeda}, R., {Rappaport}, S., {Winn}, J.~N., {et~al.} 2014,
  \href{http://dx.doi.org/10.1088/0004-637X/787/1/47}{\JournalTitle{\apj}, 787,
  47}

\bibitem[{{Sanchis-Ojeda} {et~al.}(2013){Sanchis-Ojeda}, {Rappaport}, {Winn},
  {Levine}, {Kotson}, {Latham}, \& {Buchhave}}]{SanchisOjeda+2013}
---. 2013,
  \href{http://dx.doi.org/10.1088/0004-637X/774/1/54}{\JournalTitle{\apj}, 774,
  54}

\bibitem[{{Santerne} {et~al.}(2016){Santerne}, {Moutou}, {Tsantaki}, {Bouchy},
  {H{\'e}brard}, {Adibekyan}, {Almenara}, {Amard}, {Barros}, {Boisse},
  {Bonomo}, {Bruno}, {Courcol}, {Deleuil}, {Demangeon}, {D{\'{\i}}az},
  {Guillot}, {Havel}, {Montagnier}, {Rajpurohit}, {Rey}, \&
  {Santos}}]{Santerne+2016}
{Santerne}, A., {Moutou}, C., {Tsantaki}, M., {et~al.} 2016,
  \href{http://dx.doi.org/10.1051/0004-6361/201527329}{\JournalTitle{\aap},
  587, A64}

\bibitem[{{Santos} {et~al.}(2004){Santos}, {Israelian}, \&
  {Mayor}}]{Santos+2004}
{Santos}, N.~C., {Israelian}, G., \& {Mayor}, M. 2004,
  \href{http://dx.doi.org/10.1051/0004-6361:20034469}{\JournalTitle{\aap}, 415,
  1153}

\bibitem[{{Schlaufman}(2015)}]{Schlaufman2015}
{Schlaufman}, K.~C. 2015,
  \href{http://dx.doi.org/10.1088/2041-8205/799/2/L26}{\JournalTitle{\apjl},
  799, L26}

\bibitem[{{Schlaufman} \& {Laughlin}(2011)}]{SchlaufmanLaughlin2011}
{Schlaufman}, K.~C., \& {Laughlin}, G. 2011,
  \href{http://dx.doi.org/10.1088/0004-637X/738/2/177}{\JournalTitle{\apj},
  738, 177}

\bibitem[{{Schlaufman} {et~al.}(2010){Schlaufman}, {Lin}, \&
  {Ida}}]{Schlaufman+2010}
{Schlaufman}, K.~C., {Lin}, D.~N.~C., \& {Ida}, S. 2010,
  \href{http://dx.doi.org/10.1088/2041-8205/724/1/L53}{\JournalTitle{\apjl},
  724, L53}

\bibitem[{{Sousa} {et~al.}(2011){Sousa}, {Santos}, {Israelian}, {Mayor}, \&
  {Udry}}]{Sousa+2011}
{Sousa}, S.~G., {Santos}, N.~C., {Israelian}, G., {Mayor}, M., \& {Udry}, S.
  2011,
  \href{http://dx.doi.org/10.1051/0004-6361/201117699}{\JournalTitle{\aap},
  533, A141}

\bibitem[{{Steffen} \& {Coughlin}(2016)}]{SteffenCoughlin2016}
{Steffen}, J.~H., \& {Coughlin}, J.~L. 2016,
  \href{http://dx.doi.org/10.1073/pnas.1606658113}{\JournalTitle{Proceedings of
  the National Academy of Science}, 113, 12023}

\bibitem[{{Terquem}(2014)}]{Terquem2014}
{Terquem}, C. 2014,
  \href{http://dx.doi.org/10.1093/mnras/stu1546}{\JournalTitle{\mnras}, 444,
  1738}

\bibitem[{{Twicken} {et~al.}(2016){Twicken}, {Jenkins}, {Seader}, {Tenenbaum},
  {Smith}, {Brownston}, {Burke}, {Catanzarite}, {Clarke}, {Cote}, {Girouard},
  {Klaus}, {Li}, {McCauliff}, {Morris}, {Wohler}, {Campbell}, {Kamal Uddin},
  {Zamudio}, {Sabale}, {Bryson}, {Caldwell}, {Christiansen}, {Coughlin},
  {Haas}, {Henze}, {Sanderfer}, \& {Thompson}}]{Twicken+2016}
{Twicken}, J.~D., {Jenkins}, J.~M., {Seader}, S.~E., {et~al.} 2016,
  \href{http://dx.doi.org/10.3847/0004-6256/152/6/158}{\JournalTitle{\aj}, 152,
  158}

\bibitem[{{Udry} {et~al.}(2006){Udry}, {Mayor}, {Benz}, {Bertaux}, {Bouchy},
  {Lovis}, {Mordasini}, {Pepe}, {Queloz}, \& {Sivan}}]{Udry+2006}
{Udry}, S., {Mayor}, M., {Benz}, W., {et~al.} 2006,
  \href{http://dx.doi.org/10.1051/0004-6361:20054084}{\JournalTitle{\aap}, 447,
  361}

\bibitem[{{Valsecchi} {et~al.}(2014){Valsecchi}, {Rasio}, \&
  {Steffen}}]{Valsecchi+2014}
{Valsecchi}, F., {Rasio}, F.~A., \& {Steffen}, J.~H. 2014,
  \href{http://dx.doi.org/10.1088/2041-8205/793/1/L3}{\JournalTitle{\apjl},
  793, L3}

\bibitem[{{Vanderburg} {et~al.}(2016){Vanderburg}, {Bieryla}, {Duev},
  {Jensen-Clem}, {Latham}, {Mayo}, {Baranec}, {Berlind}, {Kulkarni}, {Law},
  {Nieberding}, {Riddle}, \& {Salama}}]{Vanderburg+2016}
{Vanderburg}, A., {Bieryla}, A., {Duev}, D.~A., {et~al.} 2016,
  \href{http://dx.doi.org/10.3847/2041-8205/829/1/L9}{\JournalTitle{\apjl},
  829, L9}

\bibitem[{{Vogt} {et~al.}(1994){Vogt}, {Allen}, {Bigelow}, {Bresee}, {Brown},
  {Cantrall}, {Conrad}, {Couture}, {Delaney}, {Epps}, {Hilyard}, {Hilyard},
  {Horn}, {Jern}, {Kanto}, {Keane}, {Kibrick}, {Lewis}, {Osborne},
  {Pardeilhan}, {Pfister}, {Ricketts}, {Robinson}, {Stover}, {Tucker}, {Ward},
  \& {Wei}}]{Vogt+1994}
{Vogt}, S.~S., {Allen}, S.~L., {Bigelow}, B.~C., {et~al.} 1994,
  \href{http://dx.doi.org/10.1117/12.176725}{in \procspie, Vol. 2198,
  Instrumentation in Astronomy VIII, ed. D.~L. {Crawford} \& E.~R. {Craine}},
  362

\bibitem[{{Wang} \& {Fischer}(2015)}]{WangFischer2015}
{Wang}, J., \& {Fischer}, D.~A. 2015,
  \href{http://dx.doi.org/10.1088/0004-6256/149/1/14}{\JournalTitle{\aj}, 149,
  14}

\bibitem[{{Weiss} \& {Marcy}(2014)}]{WeissMarcy2014}
{Weiss}, L.~M., \& {Marcy}, G.~W. 2014,
  \href{http://dx.doi.org/10.1088/2041-8205/783/1/L6}{\JournalTitle{\apjl},
  783, L6}

\bibitem[{{Winn} {et~al.}(2011){Winn}, {Matthews}, {Dawson}, {Fabrycky},
  {Holman}, {Kallinger}, {Kuschnig}, {Sasselov}, {Dragomir}, {Guenther},
  {Moffat}, {Rowe}, {Rucinski}, \& {Weiss}}]{Winn+2011}
{Winn}, J.~N., {Matthews}, J.~M., {Dawson}, R.~I., {et~al.} 2011,
  \href{http://dx.doi.org/10.1088/2041-8205/737/1/L18}{\JournalTitle{\apjl},
  737, L18}

\bibitem[{{Zhu} {et~al.}(2016){Zhu}, {Wang}, \& {Huang}}]{Zhu+2016}
{Zhu}, W., {Wang}, J., \& {Huang}, C. 2016,
  \href{http://dx.doi.org/10.3847/0004-637X/832/2/196}{\JournalTitle{\apj},
  832, 196}

\end{thebibliography}

\begin{deluxetable*}{lccrrcccc}
\tabletypesize{\small}
\tablecaption{Characteristics of ``USP Planet'' sample\label{tbl:usp}}
\tablewidth{0pt}

\tablehead{
  \colhead{ID} &
  \colhead{\teff~[K]} &
  \colhead{$\log g$} &
  \colhead{[Fe/H]} &
  \colhead{$R_{\star}$ [$R_\sun$]} &
  \colhead{$M_{\star}$ [$M_\sun$]} &
  \colhead{$R_p$ [$R_\oplus$]} &
  \colhead{$P_{\rm orb}$ [hr]}
}
\renewcommand{\arraystretch}{1.0}
\startdata
KOI~0072          &       5599$^{+66}_{-65}$ &  $ 4.340^{+0.080}_{-0.100}$ & $-0.110^{+0.040}_{-0.040}$ & $ 1.060^{+0.160}_{-0.090}$ & $ 0.910^{+0.040}_{-0.030}$ & $  1.46^{+-0.17}_{--0.17}$ & 20.2\\
KOI~0191          &       5459$^{+63}_{-64}$ &  $ 4.450^{+0.070}_{-0.080}$ & $ 0.100^{+0.040}_{-0.040}$ & $ 0.940^{+0.090}_{-0.060}$ & $ 0.920^{+0.030}_{-0.030}$ & $  1.35^{+-0.12}_{--0.12}$ & 17.0\\
KOI~0577          &       5085$^{+64}_{-66}$ &  $ 4.530^{+0.040}_{-0.040}$ & $ 0.110^{+0.040}_{-0.040}$ & $ 0.820^{+0.040}_{-0.040}$ & $ 0.830^{+0.030}_{-0.030}$ & $  0.91^{+-0.09}_{--0.09}$ & 15.4\\
KOI~0717          &       5619$^{+61}_{-68}$ &  $ 4.250^{+0.090}_{-0.100}$ & $ 0.310^{+0.040}_{-0.040}$ & $ 1.280^{+0.170}_{-0.140}$ & $ 1.050^{+0.060}_{-0.040}$ & $  0.81^{+-0.11}_{--0.11}$ & 21.6\\
KOI~1128          &       5352$^{+65}_{-63}$ &  $ 4.500^{+0.060}_{-0.060}$ & $-0.040^{+0.040}_{-0.040}$ & $ 0.870^{+0.060}_{-0.050}$ & $ 0.860^{+0.030}_{-0.030}$ & $  1.24^{+-0.10}_{--0.10}$ & 23.3\\
KOI~1150          &       5755$^{+68}_{-66}$ &  $ 4.320^{+0.090}_{-0.090}$ & $ 0.100^{+0.040}_{-0.040}$ & $ 1.150^{+0.140}_{-0.110}$ & $ 1.010^{+0.040}_{-0.040}$ & $  1.02^{+-0.12}_{--0.12}$ & 16.3\\
KOI~1169          &       5634$^{+66}_{-64}$ &  $ 4.360^{+0.090}_{-0.090}$ & $ 0.110^{+0.040}_{-0.040}$ & $ 1.080^{+0.130}_{-0.100}$ & $ 0.970^{+0.030}_{-0.030}$ & $  1.52^{+-0.16}_{--0.16}$ & 16.6\\
KOI~1239          &       5747$^{+66}_{-66}$ &  $ 4.380^{+0.070}_{-0.080}$ & $-0.040^{+0.040}_{-0.040}$ & $ 1.050^{+0.110}_{-0.080}$ & $ 0.970^{+0.040}_{-0.030}$ & $  1.77^{+-0.17}_{--0.17}$ & 18.7\\
KOI~1300          &       4764$^{+63}_{-65}$ &  $ 4.580^{+0.030}_{-0.030}$ & $ 0.030^{+0.040}_{-0.040}$ & $ 0.740^{+0.020}_{-0.020}$ & $ 0.760^{+0.030}_{-0.020}$ & $  1.54^{+-0.13}_{--0.13}$ & 15.1\\
KOI~1360          &       4960$^{+64}_{-64}$ &  $ 4.590^{+0.030}_{-0.040}$ & $-0.100^{+0.040}_{-0.040}$ & $ 0.740^{+0.030}_{-0.020}$ & $ 0.780^{+0.030}_{-0.030}$ & $  0.87^{+-0.09}_{--0.09}$ & 18.2\\
KOI~1367          &       4962$^{+64}_{-64}$ &  $ 4.590^{+0.030}_{-0.040}$ & $-0.080^{+0.040}_{-0.040}$ & $ 0.750^{+0.030}_{-0.020}$ & $ 0.780^{+0.030}_{-0.030}$ & $  1.44^{+-0.12}_{--0.12}$ & 13.7\\
KOI~1428          &       4776$^{+64}_{-65}$ &  $ 4.600^{+0.030}_{-0.030}$ & $-0.110^{+0.040}_{-0.040}$ & $ 0.710^{+0.030}_{-0.020}$ & $ 0.730^{+0.030}_{-0.030}$ & $  1.90^{+-0.21}_{--0.21}$ & 22.3\\
KOI~1442          &       5568$^{+58}_{-70}$ &  $ 4.260^{+0.090}_{-0.100}$ & $ 0.390^{+0.040}_{-0.040}$ & $ 1.260^{+0.160}_{-0.140}$ & $ 1.050^{+0.060}_{-0.040}$ & $  1.43^{+-0.18}_{--0.18}$ & 16.1\\
KOI~1655          &       5536$^{+64}_{-65}$ &  $ 4.450^{+0.070}_{-0.070}$ & $-0.070^{+0.040}_{-0.040}$ & $ 0.940^{+0.080}_{-0.060}$ & $ 0.900^{+0.030}_{-0.030}$ & $  1.40^{+-0.12}_{--0.12}$ & 22.6\\
KOI~1688          &       5979$^{+71}_{-64}$ &  $ 4.100^{+0.100}_{-0.100}$ & $ 0.170^{+0.040}_{-0.040}$ & $ 1.620^{+0.290}_{-0.210}$ & $ 1.220^{+0.120}_{-0.080}$ & $  1.69^{+-0.27}_{--0.27}$ & 22.1\\
KOI~1875          &       5576$^{+65}_{-64}$ &  $ 4.400^{+0.070}_{-0.080}$ & $-0.110^{+0.040}_{-0.040}$ & $ 0.980^{+0.100}_{-0.080}$ & $ 0.900^{+0.030}_{-0.030}$ & $  1.38^{+-0.13}_{--0.13}$ & 13.0\\
KOI~2039          &       5575$^{+64}_{-64}$ &  $ 4.490^{+0.030}_{-0.060}$ & $ 0.250^{+0.040}_{-0.040}$ & $ 0.950^{+0.060}_{-0.040}$ & $ 1.010^{+0.030}_{-0.040}$ & $  0.81^{+-0.08}_{--0.08}$ & 18.2\\
KOI~2079          &       5477$^{+68}_{-66}$ &  $ 4.380^{+0.080}_{-0.090}$ & $ 0.380^{+0.040}_{-0.040}$ & $ 1.070^{+0.120}_{-0.090}$ & $ 1.000^{+0.040}_{-0.030}$ & $  0.75^{+-0.08}_{--0.08}$ & 16.6\\
KOI~2093          &       5953$^{+66}_{-65}$ &  $ 4.350^{+0.070}_{-0.080}$ & $ 0.020^{+0.040}_{-0.040}$ & $ 1.140^{+0.130}_{-0.100}$ & $ 1.070^{+0.040}_{-0.040}$ & $  1.34^{+-0.15}_{--0.15}$ & 23.8\\
KOI~2119          &       5136$^{+65}_{-65}$ &  $ 4.540^{+0.040}_{-0.050}$ & $ 0.180^{+0.040}_{-0.040}$ & $ 0.830^{+0.040}_{-0.030}$ & $ 0.870^{+0.030}_{-0.030}$ & $  1.32^{+-0.11}_{--0.11}$ & 13.7\\
KOI~2202          &       5307$^{+64}_{-65}$ &  $ 4.470^{+0.060}_{-0.070}$ & $ 0.320^{+0.040}_{-0.040}$ & $ 0.930^{+0.080}_{-0.050}$ & $ 0.940^{+0.030}_{-0.030}$ & $  1.22^{+-0.11}_{--0.11}$ & 19.4\\
KOI~2248          &       5149$^{+65}_{-64}$ &  $ 4.540^{+0.040}_{-0.050}$ & $ 0.110^{+0.040}_{-0.040}$ & $ 0.830^{+0.040}_{-0.030}$ & $ 0.850^{+0.030}_{-0.030}$ & $  1.15^{+-0.10}_{--0.10}$ & 18.2\\
KOI~2250          &       4958$^{+65}_{-66}$ &  $ 4.570^{+0.030}_{-0.040}$ & $ 0.120^{+0.040}_{-0.040}$ & $ 0.780^{+0.030}_{-0.030}$ & $ 0.820^{+0.030}_{-0.030}$ & $  1.62^{+-0.14}_{--0.14}$ & 15.1\\
KOI~2281          &       5080$^{+65}_{-65}$ &  $ 4.530^{+0.040}_{-0.040}$ & $ 0.160^{+0.040}_{-0.040}$ & $ 0.820^{+0.040}_{-0.040}$ & $ 0.840^{+0.030}_{-0.030}$ & $  0.90^{+-0.16}_{--0.16}$ & 18.5\\
KOI~2393          &       4836$^{+65}_{-65}$ &  $ 4.590^{+0.030}_{-0.040}$ & $-0.060^{+0.040}_{-0.040}$ & $ 0.730^{+0.030}_{-0.020}$ & $ 0.760^{+0.030}_{-0.030}$ & $  1.15^{+-0.10}_{--0.10}$ & 18.5\\
KOI~2396          &       5228$^{+66}_{-65}$ &  $ 4.540^{+0.040}_{-0.060}$ & $ 0.100^{+0.040}_{-0.040}$ & $ 0.830^{+0.050}_{-0.030}$ & $ 0.870^{+0.030}_{-0.030}$ & $  1.68^{+-0.17}_{--0.17}$ & 12.0\\
KOI~2409          &       4774$^{+65}_{-66}$ &  $ 4.660^{+0.030}_{-0.020}$ & $-0.590^{+0.040}_{-0.040}$ & $ 0.620^{+0.020}_{-0.020}$ & $ 0.630^{+0.020}_{-0.020}$ & $  1.26^{+-0.10}_{--0.10}$ & 13.9\\
KOI~2492          &       5635$^{+62}_{-60}$ &  $ 4.370^{+0.080}_{-0.090}$ & $-0.310^{+0.040}_{-0.040}$ & $ 0.980^{+0.110}_{-0.090}$ & $ 0.830^{+0.030}_{-0.030}$ & $  0.89^{+-0.10}_{--0.10}$ & 23.5\\
KOI~2517          &       5601$^{+64}_{-64}$ &  $ 4.520^{+0.030}_{-0.060}$ & $-0.070^{+0.040}_{-0.040}$ & $ 0.880^{+0.050}_{-0.040}$ & $ 0.940^{+0.030}_{-0.030}$ & $  1.01^{+-0.09}_{--0.09}$ & 23.3\\
KOI~2571          &       5269$^{+65}_{-66}$ &  $ 4.480^{+0.060}_{-0.070}$ & $ 0.250^{+0.040}_{-0.040}$ & $ 0.900^{+0.070}_{-0.050}$ & $ 0.910^{+0.040}_{-0.030}$ & $  1.06^{+-0.09}_{--0.09}$ & 19.9\\
KOI~2607          &       5774$^{+63}_{-65}$ &  $ 4.400^{+0.070}_{-0.090}$ & $ 0.200^{+0.040}_{-0.040}$ & $ 1.070^{+0.120}_{-0.080}$ & $ 1.050^{+0.040}_{-0.040}$ & $  1.78^{+-0.19}_{--0.19}$ & 18.0\\
KOI~2668          &       5460$^{+66}_{-66}$ &  $ 4.510^{+0.050}_{-0.060}$ & $-0.040^{+0.040}_{-0.040}$ & $ 0.870^{+0.060}_{-0.040}$ & $ 0.900^{+0.030}_{-0.030}$ & $  1.40^{+-0.12}_{--0.12}$ & 16.3\\
KOI~2694          &       4787$^{+65}_{-66}$ &  $ 4.560^{+0.030}_{-0.030}$ & $ 0.230^{+0.040}_{-0.040}$ & $ 0.770^{+0.030}_{-0.030}$ & $ 0.790^{+0.030}_{-0.030}$ & $  1.42^{+-0.12}_{--0.12}$ & 20.2\\
KOI~2753          &       5840$^{+55}_{-63}$ &  $ 4.170^{+0.100}_{-0.100}$ & $ 0.210^{+0.040}_{-0.040}$ & $ 1.450^{+0.230}_{-0.190}$ & $ 1.130^{+0.100}_{-0.060}$ & $  1.22^{+-0.18}_{--0.18}$ & 22.6\\
KOI~2756          &       5904$^{+66}_{-61}$ &  $ 4.380^{+0.070}_{-0.090}$ & $ 0.100^{+0.040}_{-0.040}$ & $ 1.100^{+0.130}_{-0.090}$ & $ 1.070^{+0.040}_{-0.040}$ & $  1.16^{+-0.12}_{--0.12}$ & 16.1\\
KOI~2763          &       4727$^{+65}_{-65}$ &  $ 4.600^{+0.030}_{-0.030}$ & $-0.010^{+0.040}_{-0.040}$ & $ 0.720^{+0.020}_{-0.020}$ & $ 0.750^{+0.030}_{-0.030}$ & $  1.14^{+-0.11}_{--0.11}$ & 12.0\\
KOI~2796          &       5686$^{+68}_{-65}$ &  $ 4.350^{+0.080}_{-0.100}$ & $ 0.000^{+0.040}_{-0.040}$ & $ 1.080^{+0.140}_{-0.100}$ & $ 0.960^{+0.030}_{-0.030}$ & $  1.09^{+-0.13}_{--0.13}$ & 13.0\\
KOI~2874          &       5243$^{+64}_{-65}$ &  $ 4.510^{+0.060}_{-0.050}$ & $-0.080^{+0.040}_{-0.040}$ & $ 0.840^{+0.050}_{-0.050}$ & $ 0.820^{+0.030}_{-0.030}$ & $  1.11^{+-0.10}_{--0.10}$ &  8.4\\
KOI~2875          &       4967$^{+63}_{-64}$ &  $ 4.580^{+0.030}_{-0.040}$ & $-0.090^{+0.040}_{-0.040}$ & $ 0.750^{+0.030}_{-0.030}$ & $ 0.770^{+0.030}_{-0.030}$ & $  1.44^{+-0.12}_{--0.12}$ &  7.2\\
KOI~2879          &       5472$^{+65}_{-66}$ &  $ 4.510^{+0.040}_{-0.060}$ & $-0.010^{+0.040}_{-0.040}$ & $ 0.880^{+0.060}_{-0.040}$ & $ 0.910^{+0.030}_{-0.030}$ & $  0.63^{+-0.05}_{--0.05}$ &  8.2\\
KOI~2916          &       4978$^{+65}_{-65}$ &  $ 4.560^{+0.040}_{-0.040}$ & $-0.000^{+0.040}_{-0.040}$ & $ 0.770^{+0.030}_{-0.030}$ & $ 0.800^{+0.030}_{-0.030}$ & $  1.00^{+-0.10}_{--0.10}$ &  7.4\\
KOI~3009          &       5110$^{+65}_{-64}$ &  $ 4.550^{+0.040}_{-0.050}$ & $ 0.170^{+0.040}_{-0.040}$ & $ 0.820^{+0.040}_{-0.030}$ & $ 0.860^{+0.030}_{-0.030}$ & $  1.04^{+-0.10}_{--0.10}$ & 18.2\\
KOI~3032          &       5213$^{+64}_{-65}$ &  $ 4.430^{+0.060}_{-0.060}$ & $ 0.360^{+0.040}_{-0.040}$ & $ 0.950^{+0.080}_{-0.060}$ & $ 0.900^{+0.030}_{-0.030}$ & $  1.43^{+-0.14}_{--0.14}$ & 15.4\\
KOI~3065          &       5713$^{+63}_{-64}$ &  $ 4.480^{+0.040}_{-0.070}$ & $-0.000^{+0.040}_{-0.040}$ & $ 0.940^{+0.070}_{-0.050}$ & $ 0.980^{+0.030}_{-0.030}$ & $  1.17^{+-0.12}_{--0.12}$ & 21.6\\
KOI~3246          &       4847$^{+66}_{-66}$ &  $ 4.580^{+0.030}_{-0.040}$ & $ 0.130^{+0.040}_{-0.040}$ & $ 0.760^{+0.030}_{-0.020}$ & $ 0.800^{+0.030}_{-0.030}$ & $  0.81^{+-0.07}_{--0.07}$ & 16.6\\
KOI~3867          &       5566$^{+67}_{-63}$ &  $ 4.420^{+0.070}_{-0.090}$ & $ 0.140^{+0.040}_{-0.040}$ & $ 0.990^{+0.100}_{-0.070}$ & $ 0.960^{+0.030}_{-0.030}$ & $  1.67^{+-0.15}_{--0.15}$ & 22.6\\
KOI~3913          &       5952$^{+65}_{-63}$ &  $ 4.260^{+0.090}_{-0.100}$ & $ 0.180^{+0.040}_{-0.040}$ & $ 1.310^{+0.190}_{-0.150}$ & $ 1.140^{+0.060}_{-0.050}$ & $  3.27^{+-0.43}_{--0.43}$ & 13.9\\
KOI~4002          &       5207$^{+64}_{-64}$ &  $ 4.530^{+0.040}_{-0.060}$ & $ 0.200^{+0.040}_{-0.040}$ & $ 0.850^{+0.050}_{-0.040}$ & $ 0.890^{+0.030}_{-0.030}$ & $  1.29^{+-0.12}_{--0.12}$ & 12.5\\
KOI~4018          &       5479$^{+64}_{-65}$ &  $ 4.520^{+0.040}_{-0.060}$ & $-0.010^{+0.040}_{-0.040}$ & $ 0.870^{+0.050}_{-0.040}$ & $ 0.920^{+0.030}_{-0.030}$ & $  1.28^{+-0.11}_{--0.11}$ & 20.9\\
KOI~4070          &       4926$^{+66}_{-65}$ &  $ 4.570^{+0.030}_{-0.040}$ & $ 0.070^{+0.040}_{-0.040}$ & $ 0.770^{+0.030}_{-0.030}$ & $ 0.800^{+0.030}_{-0.030}$ & $  1.09^{+-0.10}_{--0.10}$ & 19.0\\
KOI~4072          &       5840$^{+61}_{-65}$ &  $ 4.260^{+0.100}_{-0.100}$ & $ 0.110^{+0.040}_{-0.040}$ & $ 1.270^{+0.170}_{-0.140}$ & $ 1.060^{+0.050}_{-0.040}$ & $  1.02^{+-0.13}_{--0.13}$ & 16.6\\
KOI~4109          &       4995$^{+66}_{-66}$ &  $ 4.530^{+0.040}_{-0.040}$ & $ 0.250^{+0.040}_{-0.040}$ & $ 0.830^{+0.040}_{-0.040}$ & $ 0.840^{+0.030}_{-0.030}$ & $  0.72^{+-0.07}_{--0.07}$ & 15.8\\
KOI~4144          &       6000$^{+65}_{-64}$ &  $ 4.390^{+0.070}_{-0.080}$ & $-0.110^{+0.040}_{-0.040}$ & $ 1.080^{+0.110}_{-0.080}$ & $ 1.030^{+0.040}_{-0.040}$ & $  1.19^{+-0.12}_{--0.12}$ & 23.5\\
KOI~4159          &       5233$^{+66}_{-66}$ &  $ 4.490^{+0.060}_{-0.050}$ & $ 0.120^{+0.040}_{-0.040}$ & $ 0.880^{+0.050}_{-0.050}$ & $ 0.860^{+0.030}_{-0.030}$ & $  0.75^{+-0.07}_{--0.07}$ & 23.3\\
KOI~4199          &       5109$^{+65}_{-65}$ &  $ 4.570^{+0.030}_{-0.050}$ & $-0.120^{+0.040}_{-0.040}$ & $ 0.770^{+0.040}_{-0.030}$ & $ 0.800^{+0.030}_{-0.030}$ & $  0.77^{+-0.07}_{--0.07}$ & 13.0\\
KOI~4366          &       5269$^{+66}_{-63}$ &  $ 4.530^{+0.050}_{-0.060}$ & $-0.110^{+0.040}_{-0.040}$ & $ 0.820^{+0.050}_{-0.040}$ & $ 0.830^{+0.030}_{-0.030}$ & $  1.23^{+-0.12}_{--0.12}$ & 18.2\\
KOI~4430          &       5104$^{+65}_{-66}$ &  $ 4.550^{+0.040}_{-0.050}$ & $ 0.100^{+0.040}_{-0.040}$ & $ 0.810^{+0.040}_{-0.030}$ & $ 0.840^{+0.030}_{-0.030}$ & $  1.38^{+-0.20}_{--0.20}$ & 12.2\\
KOI~4441          &       4888$^{+65}_{-65}$ &  $ 4.570^{+0.030}_{-0.030}$ & $ 0.060^{+0.040}_{-0.040}$ & $ 0.760^{+0.030}_{-0.030}$ & $ 0.790^{+0.030}_{-0.030}$ & $  1.41^{+-0.18}_{--0.18}$ & 16.3\\
KOI~4469          &       4909$^{+65}_{-64}$ &  $ 4.570^{+0.030}_{-0.040}$ & $ 0.070^{+0.040}_{-0.040}$ & $ 0.770^{+0.030}_{-0.030}$ & $ 0.800^{+0.030}_{-0.030}$ & $  0.71^{+-0.07}_{--0.07}$ & 21.4\\
KOI~4746          &       4948$^{+66}_{-66}$ &  $ 4.570^{+0.030}_{-0.040}$ & $ 0.080^{+0.040}_{-0.040}$ & $ 0.780^{+0.030}_{-0.030}$ & $ 0.810^{+0.030}_{-0.030}$ & $  0.83^{+-0.08}_{--0.08}$ & 23.5\\
KOI~4841          &       4803$^{+65}_{-65}$ &  $ 4.590^{+0.030}_{-0.030}$ & $-0.010^{+0.040}_{-0.040}$ & $ 0.730^{+0.030}_{-0.020}$ & $ 0.760^{+0.030}_{-0.030}$ & $  1.35^{+-0.13}_{--0.13}$ & 17.0\\
KIC~8435766      &       5060$^{+65}_{-64}$ &  $ 4.570^{+0.030}_{-0.040}$ & $ 0.000^{+0.040}_{-0.040}$ & $ 0.780^{+0.030}_{-0.030}$ & $ 0.820^{+0.030}_{-0.030}$ & $  1.25^{+ 0.14}_{- 0.14}$ &  8.5\\
KIC~11187332     &       5573$^{+66}_{-65}$ &  $ 4.430^{+0.070}_{-0.070}$ & $-0.090^{+0.040}_{-0.040}$ & $ 0.960^{+0.090}_{-0.070}$ & $ 0.900^{+0.030}_{-0.030}$ & $  1.17^{+ 0.17}_{- 0.17}$ &  7.3\\
KIC~2718885      &       5614$^{+64}_{-64}$ &  $ 4.370^{+0.080}_{-0.090}$ & $ 0.120^{+0.040}_{-0.040}$ & $ 1.060^{+0.120}_{-0.090}$ & $ 0.960^{+0.030}_{-0.030}$ & $  1.12^{+ 0.19}_{- 0.19}$ &  4.7
\enddata
\end{deluxetable*}

\begin{deluxetable*}{lccrrcccc}
\tabletypesize{\small}
\tablecaption{Characteristics of ``Hot Jupiter'' sample\label{tbl:giant}}
\tablewidth{0pt}

\tablehead{
  \colhead{ID} &
  \colhead{\teff~[K]} &
  \colhead{$\log g$} &
  \colhead{[Fe/H]} &
  \colhead{$R_{\star}$ [$R_\sun$]} &
  \colhead{$M_{\star}$ [$M_\sun$]} &
  \colhead{$R_p$ [$R_\oplus$]} &
  \colhead{$P_{\rm orb}$ [hr]}
}
\renewcommand{\arraystretch}{1.0}
\startdata
KOI~0001          &       5815$^{+66}_{-65}$ &  $ 4.390^{+0.080}_{-0.090}$ & $ 0.010^{+0.040}_{-0.040}$ & $ 1.060^{+0.120}_{-0.090}$ & $ 1.010^{+0.030}_{-0.030}$ & $ 14.32^{+-1.42}_{--1.42}$ & 59.3\\
KOI~0003          &       4867$^{+66}_{-65}$ &  $ 4.540^{+0.040}_{-0.030}$ & $ 0.310^{+0.040}_{-0.040}$ & $ 0.810^{+0.030}_{-0.030}$ & $ 0.830^{+0.030}_{-0.030}$ & $  5.11^{+-0.41}_{--0.41}$ &117.4\\
KOI~0007          &       5833$^{+60}_{-67}$ &  $ 4.120^{+0.110}_{-0.100}$ & $ 0.170^{+0.040}_{-0.040}$ & $ 1.530^{+0.240}_{-0.200}$ & $ 1.120^{+0.100}_{-0.060}$ & $  4.13^{+-0.60}_{--0.60}$ & 77.0\\
KOI~0017          &       5667$^{+58}_{-63}$ &  $ 4.170^{+0.100}_{-0.100}$ & $ 0.340^{+0.040}_{-0.040}$ & $ 1.450^{+0.220}_{-0.180}$ & $ 1.110^{+0.100}_{-0.060}$ & $ 15.04^{+-2.10}_{--2.10}$ & 77.5\\
KOI~0022          &       5885$^{+61}_{-58}$ &  $ 4.210^{+0.100}_{-0.100}$ & $ 0.200^{+0.040}_{-0.040}$ & $ 1.380^{+0.210}_{-0.180}$ & $ 1.130^{+0.090}_{-0.060}$ & $ 14.20^{+-2.02}_{--2.02}$ &189.4\\
KOI~0063          &       5660$^{+64}_{-63}$ &  $ 4.490^{+0.030}_{-0.050}$ & $ 0.230^{+0.040}_{-0.040}$ & $ 0.960^{+0.050}_{-0.040}$ & $ 1.030^{+0.030}_{-0.030}$ & $  6.09^{+-0.49}_{--0.49}$ &226.3\\
KOI~0127          &       5600$^{+58}_{-67}$ &  $ 4.350^{+0.080}_{-0.090}$ & $ 0.350^{+0.040}_{-0.040}$ & $ 1.130^{+0.140}_{-0.100}$ & $ 1.040^{+0.040}_{-0.040}$ & $ 12.10^{+-1.29}_{--1.29}$ & 85.9\\
KOI~0128          &       5669$^{+66}_{-67}$ &  $ 4.210^{+0.090}_{-0.100}$ & $ 0.250^{+0.040}_{-0.040}$ & $ 1.330^{+0.210}_{-0.150}$ & $ 1.050^{+0.070}_{-0.050}$ & $ 14.65^{+-1.99}_{--1.99}$ &118.6\\
KOI~0135          &       5951$^{+68}_{-66}$ &  $ 4.210^{+0.100}_{-0.120}$ & $ 0.320^{+0.040}_{-0.040}$ & $ 1.450^{+0.260}_{-0.190}$ & $ 1.230^{+0.090}_{-0.060}$ & $ 13.03^{+-2.05}_{--2.05}$ & 72.5\\
KOI~0141          &       5322$^{+65}_{-63}$ &  $ 4.430^{+0.070}_{-0.070}$ & $ 0.300^{+0.040}_{-0.040}$ & $ 0.970^{+0.090}_{-0.070}$ & $ 0.930^{+0.030}_{-0.030}$ & $  5.68^{+-0.53}_{--0.53}$ & 62.9\\
KOI~0186          &       5802$^{+62}_{-63}$ &  $ 4.350^{+0.080}_{-0.090}$ & $ 0.180^{+0.040}_{-0.040}$ & $ 1.130^{+0.140}_{-0.100}$ & $ 1.050^{+0.040}_{-0.040}$ & $ 14.97^{+-1.60}_{--1.60}$ & 77.8\\
KOI~0195          &       5535$^{+64}_{-66}$ &  $ 4.480^{+0.060}_{-0.070}$ & $-0.160^{+0.040}_{-0.040}$ & $ 0.890^{+0.080}_{-0.060}$ & $ 0.870^{+0.030}_{-0.030}$ & $ 11.56^{+-0.93}_{--0.93}$ & 77.3\\
KOI~0201          &       5526$^{+69}_{-67}$ &  $ 4.240^{+0.100}_{-0.100}$ & $ 0.350^{+0.040}_{-0.040}$ & $ 1.260^{+0.170}_{-0.140}$ & $ 1.020^{+0.050}_{-0.040}$ & $ 10.93^{+-1.35}_{--1.35}$ &101.5\\
KOI~0203          &       5714$^{+64}_{-64}$ &  $ 4.440^{+0.040}_{-0.070}$ & $ 0.310^{+0.040}_{-0.040}$ & $ 1.030^{+0.080}_{-0.050}$ & $ 1.080^{+0.030}_{-0.040}$ & $ 15.05^{+-1.20}_{--1.20}$ & 35.8\\
KOI~0214          &       5481$^{+72}_{-66}$ &  $ 4.310^{+0.090}_{-0.090}$ & $ 0.390^{+0.040}_{-0.040}$ & $ 1.160^{+0.140}_{-0.120}$ & $ 1.000^{+0.040}_{-0.040}$ & $ 11.19^{+-1.27}_{--1.27}$ & 79.4\\
KOI~0439          &       5458$^{+65}_{-65}$ &  $ 4.370^{+0.080}_{-0.090}$ & $ 0.320^{+0.040}_{-0.040}$ & $ 1.070^{+0.120}_{-0.090}$ & $ 0.980^{+0.030}_{-0.030}$ & $  5.05^{+-0.51}_{--0.51}$ & 45.6\\
KOI~0466          &       5954$^{+66}_{-63}$ &  $ 4.210^{+0.100}_{-0.100}$ & $ 0.040^{+0.040}_{-0.040}$ & $ 1.360^{+0.190}_{-0.160}$ & $ 1.090^{+0.050}_{-0.040}$ & $ 10.66^{+-1.79}_{--1.79}$ &225.4\\
KOI~0760          &       5741$^{+66}_{-65}$ &  $ 4.360^{+0.080}_{-0.090}$ & $ 0.090^{+0.040}_{-0.040}$ & $ 1.090^{+0.130}_{-0.100}$ & $ 1.000^{+0.040}_{-0.030}$ & $ 12.66^{+-1.31}_{--1.31}$ &119.0\\
KOI~0800          &       5904$^{+63}_{-61}$ &  $ 4.250^{+0.100}_{-0.100}$ & $ 0.200^{+0.040}_{-0.040}$ & $ 1.310^{+0.200}_{-0.150}$ & $ 1.130^{+0.070}_{-0.050}$ & $  4.45^{+-0.63}_{--0.63}$ & 65.0\\
KOI~0889          &       5311$^{+63}_{-66}$ &  $ 4.480^{+0.060}_{-0.070}$ & $ 0.220^{+0.040}_{-0.040}$ & $ 0.910^{+0.070}_{-0.050}$ & $ 0.900^{+0.040}_{-0.030}$ & $ 11.86^{+-0.95}_{--0.95}$ &213.1\\
KOI~1779          &       5861$^{+63}_{-65}$ &  $ 4.420^{+0.040}_{-0.060}$ & $ 0.300^{+0.040}_{-0.040}$ & $ 1.080^{+0.080}_{-0.060}$ & $ 1.130^{+0.040}_{-0.040}$ & $  4.35^{+-0.35}_{--0.35}$ &111.8\\
KOI~1800          &       5611$^{+65}_{-65}$ &  $ 4.510^{+0.030}_{-0.050}$ & $ 0.070^{+0.040}_{-0.040}$ & $ 0.910^{+0.050}_{-0.030}$ & $ 0.980^{+0.030}_{-0.030}$ & $  6.29^{+-0.55}_{--0.55}$ &187.0\\
KOI~3689          &       5988$^{+68}_{-67}$ &  $ 4.190^{+0.100}_{-0.110}$ & $ 0.030^{+0.040}_{-0.040}$ & $ 1.410^{+0.220}_{-0.170}$ & $ 1.110^{+0.060}_{-0.040}$ & $ 14.08^{+-1.98}_{--1.98}$ &125.8
\enddata
\end{deluxetable*}

\normalsize
\begin{deluxetable*}{lcccc}
\tabletypesize{\small}
\tablecaption{Comparisons between metallicity distributions\label{tbl:pvalues}}
\tablewidth{0pt}
\tablehead{
  \colhead{Sample} & \colhead{Number} & \colhead{Sample mean} &  \multicolumn{2}{c}{$p$ for comparison\tablenotemark{a} with} \\
  \colhead{name}   & \colhead{of stars}  & \colhead{[Fe/H]}   &  \colhead{USPs} & \colhead{Hot Jupiters} 
  }
\renewcommand{\arraystretch}{1.00}
\startdata
USP Planets       & 64  & $0.0584 \pm 0.0050$ & $\cdots$            & $3\times 10^{-4}$ \\
Hot Jupiters      & 23  & $0.2096 \pm 0.0085$ & $3\times 10^{-4}$  & $\cdots$ \\
Hot Small Planets & 246 & $0.0459 \pm 0.0026$ & $0.39$              & $2\times 10^{-5}$
\enddata
\tablenotetext{a}{Probability of being drawn from the same distribution, based on a two-sample Kolmogorov-Smirnov test.}  
\end{deluxetable*}
\renewcommand{\arraystretch}{1.00}

\end{document}